\documentclass[reprint,amsmath,amssymb,aip,jcp]{revtex4-1}

\usepackage{graphicx}
\usepackage{dcolumn}
\usepackage{bm}
\usepackage{amsfonts}
\usepackage{color}
\usepackage{amsbsy}

\newcommand\beq{\begin{equation}}
\newcommand\eeq{\end{equation}}
\newcommand\beqa{\begin{eqnarray}}
\newcommand\eeqa{\end{eqnarray}}

\newcommand{\co}{\text{coex}}
\newcommand{\CS}{\text{CS}}
\newcommand{\CSK}{\text{CSK}}
\newcommand{\BP}{\text{BP}}
\newcommand{\RV}{\text{RV}}

\begin{document}



\title{Note: Equation of state and the freezing point in the hard-sphere model}

\author{Miguel Robles}
\email{mrp@ier.unam.mx}
\homepage{http://xml.ier.unam.mx/xml/tc/ft/mrp/}
\affiliation{Instituto de Energ\'{\i}as Renovables, Universidad Nacional Aut\'onoma de M\'exico (U.N.A.M.),
Temixco, Morelos 62580, M{e}xico}

\author{Mariano L\'{o}pez de Haro}
\email{malopez@unam.mx}
\homepage{http://xml.ier.unam.mx/xml/tc/ft/mlh/}
\affiliation{Instituto de Energ\'{\i}as Renovables, Universidad Nacional Aut\'onoma de M\'exico (U.N.A.M.),
Temixco, Morelos 62580, M{e}xico}
\author{Andr\'es Santos}
\email{andres@unex.es}
\homepage{http://www.unex.es/eweb/fisteor/andres/}
\affiliation{Departamento de F\'{\i}sica, Universidad de
Extremadura, Badajoz, E-06071, Spain}

\date{\today}

\begin{abstract}
The merits of different analytical equations of state for the hard-sphere system with respect to the recently computed high-accuracy value of the freezing-point packing fraction are assessed. It is found that the Carnahan--Starling--Kolafa and the branch-point approximant equations of state yield the best performance.
\end{abstract}

\maketitle

{Despite the simplicity of the hard-sphere (HS) intermolecular potential and the vast amount of studies devoted to this model, up to date no one has been able to derive analytically neither the free energy nor the phase diagram of the HS system. Therefore, many of the important results concerning the equilibrium properties of the HS model have been obtained from computer simulations.}
It is well known that in the HS system the absolute temperature $T$ only enters as a scaling parameter and so its {equation of state (EOS)} is usually presented as a graph in the  compressibility factor ($Z\equiv {p}/{\rho k_B T}$, {with $p$, $\rho$, and $k_B$ being the pressure, number density, and Boltzmann constant, respectively}) vs packing fraction ($\eta\equiv\frac{\pi}{6}\rho\sigma^3$, $\sigma$ being the diameter of the spheres) plane.\cite{M08}
The characteristics of this {diagram}  are relatively well understood, at least qualitatively. It comprises a stable fluid branch going from $\eta=0$ to the freezing packing fraction $\eta_\text{f}\simeq 0.492$, where a fluid-solid phase transition takes place,\cite{AW57,FMSV12} a region of fluid-solid  coexistence from $\eta_\text{f}$ to the the crystal melting point  $\eta_\text{m}\simeq 0.543$,\cite{HR68,FMSV12}  and finally a stable solid (crystalline) branch from $\eta_\text{m}$ to the close-packing fraction {$\eta_{\text{cp}}=\frac{\pi}{6}\sqrt{2}\simeq 0.7405$.\cite{S98b}} Beyond the freezing point there is also a region of metastable fluid states that is supposed to end at the packing fraction $\eta_\text{g} \simeq 0.58$,\cite{S98b,PZ10} where a widely accepted glass transition  {occurs}. The glass branch ends at
$\eta_{\text{rcp}}\simeq 0.64$ corresponding to the random close-packing of an {amorphous solid.\cite{BM60}}  There is further a region of metastable crystalline states for packing fractions below $\eta_\text{m}$.

Recently, accurate tethered Monte Carlo (MC) simulations have been reported\cite{FMSV12} in which the fluid-solid coexistence pressure ($p_{\co}$) of the HS system was computed, namely {$p_{\co}^*\equiv (\sigma^3/k_BT) p_\co=11.5727(10)$}, the number enclosed by parentheses denoting the statistical error. The specific volumes associated with the freezing and melting points were also reported with the values $v_\text{f}=1/\rho_{\text{f}}=1.06448(10)\sigma^3$ and $v_\text{m}=1/\rho_{\text{m}}=0.96405(3)\sigma^3$, respectively.

Given these  results, the aim of this Note is to explore whether starting with the above high-accuracy estimate of $p_{\co}$ and determining the freezing-point packing fraction (with its associated statistical error) from available analytical EOS one may conclude which one yields the best performance near the freezing point.
To achieve our goal, we will examine the following four analytical EOS. First, we recall the celebrated
Carnahan--Starling (CS)\cite{CS69} EOS:
\begin{equation}
Z_{\CS}=\frac{1+\eta+\eta ^2-\eta ^3 }{(1-\eta )^3}.
\label{CS}
\end{equation}
Next, we consider Kolafa's correction, i.e., the
Carnahan--Starling--Kolafa (CSK)\cite{CSK86} EOS:
\begin{equation}
Z_{\CSK}=\frac{1+\eta+\eta ^2 -\frac{2}{3} (1+\eta ) \eta ^3}{(1-\eta )^3}.
\label{CSK}
\end{equation}
As a third EOS, a proposal based on the so-called rescaled virial (RV) expansion\cite{BC87} will also be included, namely
\begin{equation}
Z_{\RV}=\frac{1+\sum_{n=1}^6  C_n \eta^n}{(1-\eta)^3},
\label{RV}
\end{equation}
with $C_1=C_2=1$, and {$C_n=\sum_{j=0}^3 \binom{3}{j}(-1)^{j+1} b_{n-2+j}$ for $n=3$--$6$}, $b_j$ being the (reduced) virial coefficients. Finally, a recently proposed
branch-point (BP) approximant\cite{SH09} will be considered. It reads
\begin{equation}
Z_{\BP}=1+\frac{1+\sum_{n=1}^3c_n \eta^n-(1+2a_1 \eta+a_2 \eta^2)^{3/2}}{A(1-\eta )^3},
\label{BP}
\end{equation}
with $a_1=-C_5/C_4$, $a_2=7a_1^2-6C_6/C_4$, $A=-\frac{3}{8}(a_2-a_1^2)^2/C_4$, $c_1=3a_1+4A$, $c_2=\frac{3}{2}(a_2+a_1^2)-2A$, and $c_3=\frac{1}{2}a_1(3a_2-a_1^2)+(b_4-18)A$.
{One should add in connection with Eqs.\ (\ref{RV}) and (\ref{BP}) that  they require the first seven virial coefficients. Only  $b_2=4$, $b_3=10$, and $b_4=\frac{219\sqrt{2}-712\pi+4131\tan^{-1}\sqrt{2}}{35\pi}$ are  exactly known, while  $b_5=28.22445(10)$, $b_6=39.81550(36)$, and $b_7=53.3413(16)$ have been determined numerically.\cite{LKM05}}

The procedure involves inverting Eqs.\ (\ref{CS})--(\ref{BP}) to compute $\eta_\text{f}$ (and its statistical error $\Delta\eta_\text{f}$) from the MC value of $p_{\co}^*$ (and its associated statistical error $\Delta p_{\co}^*=10^{-3}$). The four EOS give $\left.\partial p^*/\partial\eta\right|_{\eta=0.492}\approx 100$--$101$, so
that one can easily estimate $\Delta\eta_\text{f}\approx 10^{-5}$.
{However, although the numerical inversion of Eqs.\ \eqref{CS} and \eqref{CSK} is straightforward, there are complications associated with Eqs.\ \eqref{RV} and \eqref{BP} due to the statistical uncertainties on the higher order virial coefficients. To take these into account we used the following procedure.}
(i) A random number $p^*$ is generated having a normal distribution with average value $p_{\co}^*$ and standard deviation  $\Delta p_{\co}^*=10^{-3}$; (ii) a value of the packing fraction $\eta$ is derived  through the  equation
$\frac{6}{\pi} \eta Z(\eta)=p^*$,
where $Z(\eta)$ is the compressibility factor corresponding to each one of the above EOS; (iii)
step (i) is repeated so as to gather a statistically representative set of $\mathcal{N}$ values of $\eta$; and (iv)
finally, $\eta_\text{f}$ is taken as the average of the above solutions and the standard deviation $\Delta\eta_{\text{f}}$ is equated to the associated statistical error.
{In the cases of Eqs.\ \eqref{RV} and \eqref{BP} we also accounted for the statistical errors associated with $b_5$--$b_7$ in the MC procedure, but we observed that their influence was practically negligible. The number of elements were chosen as $\mathcal{N}=5\times 10^4$ for Eqs.\ \eqref{CS} and \eqref{CSK} and $\mathcal{N}=1.5\times 10^5$  for Eqs.\ \eqref{RV} and \eqref{BP}.}

 \begin{table}
\caption{{Freezing-point packing fraction $\eta_{\text{f}}$ as measured in tethered MC simulations\cite{FMSV12} and as derived  from Eqs.\  (\protect\ref{CS})--(\protect\ref{BP})   and from a fit to MD simulation data.\cite{BLW10} The third column provides the  excess chemical potential at the freezing point, $\beta\mu_{\text{f}}^{\text{ex}}$, as derived from Eqs.\ (\protect\ref{CS})--(\protect\ref{BP})   and from a  fit to MC simulation data.\cite{LS94}}}\label{tab1}
\begin{ruledtabular}
\begin{tabular}{ccc}
Method&$\eta_{\text{f}}$&${\beta\mu_{\text{f}}^{\text{ex}}}$ \\  \hline
Tethered MC\footnote{Reference \protect\onlinecite{FMSV12}}&$0.491882(46)$& $\cdots$\\
CS&  $0.491972(10)$&{$16.1119(11)$} \\
CSK& $0.491927(10)$	&{$16.1395(10)$}\\
RV&${0.491820}(10)$&{$16.1404(11)$}\\
BP&${0.491917}(10)$&{$16.1289(11)$}\\
MD\footnote{Reference \protect\onlinecite{BLW10}}
and MC\footnote{Reference \protect\onlinecite{LS94}}
&$0.491835(11)$&{$16.167(54)$}\\
     \end{tabular}
 \end{ruledtabular}
 \end{table}

After applying the previous procedure to each one of the EOS (\ref{CS})--(\ref{BP}), the results shown in {the second column of} Table \ref{tab1} were obtained. The simulation value of $\eta_\text{f}$ that follows from the value of the freezing-point specific volume $v_\text{f}$ stated earlier is also included in Table \ref{tab1}.
{Additionally, Table \ref{tab1} contains an estimate of $\eta_\text{f}$ obtained by application of the procedure outlined above to a quadratic fit  to  recent rather accurate  molecular dynamics (MD) simulation data, together with their error bars,\cite{BLW10}  for the three closest densities ($\rho\sigma^3=0.930$, $0.940$, and $0.950$) to the freezing density.}

\begin{figure}
\begin{center}
\includegraphics[width=7.5cm]{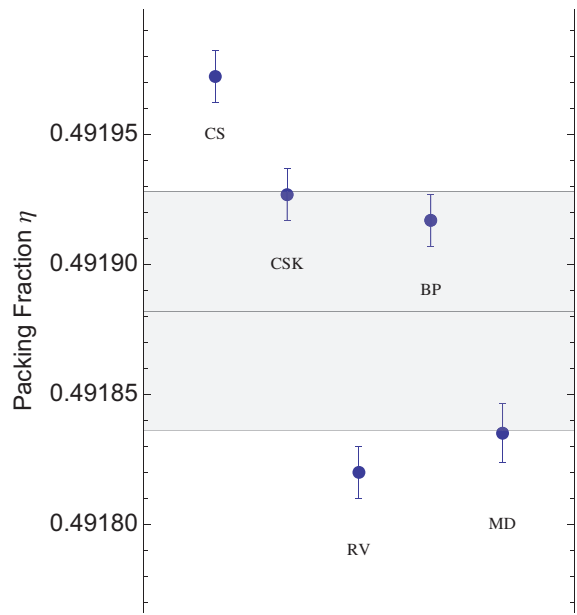}
\caption{Values of the HS freezing-point packing fraction $\eta_{\text{f}}$, together with their error bars, as obtained from Eqs.\  (\protect\ref{CS})--(\protect\ref{BP}) {and from the quadratic fit to MD data}.\cite{BLW10} The shaded area represents the error bar corresponding to the simulation result of Ref.\ \protect\onlinecite{FMSV12}.
\label{F1}}
\end{center}
\end{figure}

The results of Table \ref{tab1} for $\eta_\text{f}$ are graphically displayed in Fig.\ \ref{F1}.
It is clear that the best performance with respect to the simulation results is provided by both $Z_{\CSK}$ and $Z_{\BP}$, with possibly a slight superiority of the latter. Whereas $Z_{\CSK}$ is simpler than $Z_{\BP}$, the latter has the advantage of predicting a physical value (smaller than  $\eta_{\text{cp}}$) for the radius of convergence of the virial series.\cite{SH09,CM06}
{It is also interesting to note that the MD estimate and the MC value of $\eta_\text{f}$ are statistically consistent since the difference  between them is slightly smaller than the combined standard deviation.}

{It might be argued that using a single density--pressure point at freezing  is not sufficient for a fair assessment of the whole stable fluid
branch. To account for this, we have also analyzed the excess chemical potential at freezing, $\beta\mu_\text{f}^{\text{ex}}=Z(\eta_\text{f})-1+
\int_0^{\eta_\text{f}}d\eta'[Z(\eta')-1]/\eta'$, which requires integration over the whole fluid range. We have evaluated  $\beta\mu_\text{f}^{\text{ex}}$ from Eqs.\ \eqref{CS}--\eqref{BP} by following a procedure similar to the one described above (with $\mathcal{N}=1.5\times 10^5$), except that the values of $\eta_\text{f}$ along with their uncertainties are now used. The results are displayed in the third column of Table \ref{tab1}. Since $\beta\mu_\text{f}^{\text{ex}}$ was not directly reported in Ref.\ \onlinecite{FMSV12}, we have resorted to MC results of Ref.\ \onlinecite{LS94} for $\rho\sigma^3=0.925$, $0.94$, and $1.0$ and applied our procedure (again with $\mathcal{N}=1.5\times 10^5$) to a quadratic fit. Except in the CS case, the theoretical values deviate from the MC estimate less than the combined standard deviation. In any case, a  more accurate simulation value for $\beta\mu_\text{f}^{\text{ex}}$ would be needed to discriminate among the CSK, RV, and BP  predictions.}

Two of us (A.S. and M.L.H.) acknowledge the financial support of the Spanish Government through Grant No.\ FIS2010-16587 and  the Junta de Extremadura (Spain) through Grant No.\ GR10158 (partially financed by FEDER funds).



\begin{thebibliography}{15}
\expandafter\ifx\csname natexlab\endcsname\relax\def\natexlab#1{#1}\fi
\expandafter\ifx\csname bibnamefont\endcsname\relax
  \def\bibnamefont#1{#1}\fi
\expandafter\ifx\csname bibfnamefont\endcsname\relax
  \def\bibfnamefont#1{#1}\fi
\expandafter\ifx\csname citenamefont\endcsname\relax
  \def\citenamefont#1{#1}\fi
\expandafter\ifx\csname url\endcsname\relax
  \def\url#1{\texttt{#1}}\fi
\expandafter\ifx\csname urlprefix\endcsname\relax\def\urlprefix{URL }\fi
\providecommand{\bibinfo}[2]{#2}
\providecommand{\eprint}[2][]{\url{#2}}

\bibitem[{\citenamefont{Mulero}(2008)}]{M08}
\bibinfo{editor}{\bibfnamefont{A.}~\bibnamefont{Mulero}}, ed.,
  \emph{\bibinfo{title}{Theory and Simulation of Hard-Sphere Fluids and Related
  Systems}} (\bibinfo{publisher}{Springer-Verlag}, \bibinfo{address}{Berlin},
  \bibinfo{year}{2008}), vol. \bibinfo{volume}{753} of
  \emph{\bibinfo{series}{Lectures Notes in Physics}}.

\bibitem[{\citenamefont{Alder and Wainwright}(1957)}]{AW57}
\bibinfo{author}{\bibfnamefont{B.~J.} \bibnamefont{Alder}} \bibnamefont{and}
  \bibinfo{author}{\bibfnamefont{T.~E.} \bibnamefont{Wainwright}},
  \bibinfo{journal}{J. Chem. Phys.} \textbf{\bibinfo{volume}{27}},
  \bibinfo{pages}{1208} (\bibinfo{year}{1957}).

\bibitem[{\citenamefont{Fern\'andez et~al.}(2012)\citenamefont{Fern\'andez,
  Mart\'in-Mayor, Seoane, and Verrocchio}}]{FMSV12}
\bibinfo{author}{\bibfnamefont{L.~A.} \bibnamefont{Fern\'andez}},
  \bibinfo{author}{\bibfnamefont{V.}~\bibnamefont{Mart\'in-Mayor}},
  \bibinfo{author}{\bibfnamefont{B.}~\bibnamefont{Seoane}}, \bibnamefont{and}
  \bibinfo{author}{\bibfnamefont{P.}~\bibnamefont{Verrocchio}},
  \bibinfo{journal}{Phys. Rev. Lett.} \textbf{\bibinfo{volume}{108}},
  \bibinfo{pages}{165701} (\bibinfo{year}{2012}).

\bibitem[{\citenamefont{Hoover and Ree}(1968)}]{HR68}
\bibinfo{author}{\bibfnamefont{W.~G.} \bibnamefont{Hoover}} \bibnamefont{and}
  \bibinfo{author}{\bibfnamefont{F.~H.} \bibnamefont{Ree}},
  \bibinfo{journal}{J. Chem. Phys.} \textbf{\bibinfo{volume}{49}},
  \bibinfo{pages}{3609} (\bibinfo{year}{1968}).

\bibitem[{\citenamefont{Speedy}(1998)}]{S98b}
\bibinfo{author}{\bibfnamefont{R.~J.} \bibnamefont{Speedy}},
  \bibinfo{journal}{Mol. Phys.} \textbf{\bibinfo{volume}{95}},
  \bibinfo{pages}{169} (\bibinfo{year}{1998}).

\bibitem[{\citenamefont{Parisi and Zamponi}(2010)}]{PZ10}
\bibinfo{author}{\bibfnamefont{G.}~\bibnamefont{Parisi}} \bibnamefont{and}
  \bibinfo{author}{\bibfnamefont{F.}~\bibnamefont{Zamponi}},
  \bibinfo{journal}{Rev. Mod. Phys.} \textbf{\bibinfo{volume}{82}},
  \bibinfo{pages}{789} (\bibinfo{year}{2010}).

\bibitem[{\citenamefont{Bernal and Mason}(1960)}]{BM60}
\bibinfo{author}{\bibfnamefont{J.}~\bibnamefont{Bernal}} \bibnamefont{and}
  \bibinfo{author}{\bibfnamefont{J.}~\bibnamefont{Mason}},
  \bibinfo{journal}{Nature} \textbf{\bibinfo{volume}{188}},
  \bibinfo{pages}{910} (\bibinfo{year}{1960}).

\bibitem[{\citenamefont{Carnahan and Starling}(1969)}]{CS69}
\bibinfo{author}{\bibfnamefont{N.~F.} \bibnamefont{Carnahan}} \bibnamefont{and}
  \bibinfo{author}{\bibfnamefont{K.~E.} \bibnamefont{Starling}},
  \bibinfo{journal}{J. Chem. Phys.} \textbf{\bibinfo{volume}{51}},
  \bibinfo{pages}{635} (\bibinfo{year}{1969}).

\bibitem[{CSK()}]{CSK86}
\bibinfo{note}{This EOS is a slight modification by J.\ Kolafa of the CS EOS.
  It first appeared as Eq.\ (4.46) in the review paper by T.\ Boubl\'{i}k and
  I.\ Nezbeda, Collect.\ Czech.\ Chem.\ Commun.\ \textbf{51}, 2301 (1986).}

\bibitem[{\citenamefont{Baus and Colot}(1987)}]{BC87}
\bibinfo{author}{\bibfnamefont{M.}~\bibnamefont{Baus}} \bibnamefont{and}
  \bibinfo{author}{\bibfnamefont{J.~L.} \bibnamefont{Colot}},
  \bibinfo{journal}{Phys. Rev. A} \textbf{\bibinfo{volume}{36}},
  \bibinfo{pages}{3912} (\bibinfo{year}{1987}).

\bibitem[{\citenamefont{Santos and { L\'opez de Haro}}(2009)}]{SH09}
\bibinfo{author}{\bibfnamefont{A.}~\bibnamefont{Santos}} \bibnamefont{and}
  \bibinfo{author}{\bibfnamefont{M.}~\bibnamefont{{ L\'opez de Haro}}},
  \bibinfo{journal}{J. Chem. Phys.} \textbf{\bibinfo{volume}{130}},
  \bibinfo{pages}{214104} (\bibinfo{year}{2009}).

\bibitem[{\citenamefont{Lab\'ik et~al.}(2005)\citenamefont{Lab\'ik, Kolafa, and
  Malijevsk\'y}}]{LKM05}
\bibinfo{author}{\bibfnamefont{S.}~\bibnamefont{Lab\'ik}},
  \bibinfo{author}{\bibfnamefont{J.}~\bibnamefont{Kolafa}}, \bibnamefont{and}
  \bibinfo{author}{\bibfnamefont{A.}~\bibnamefont{Malijevsk\'y}},
  \bibinfo{journal}{Phys. Rev. E} \textbf{\bibinfo{volume}{71}},
  \bibinfo{pages}{021105} (\bibinfo{year}{2005}).

\bibitem[{\citenamefont{Bannerman et~al.}(2010)\citenamefont{Bannerman, Lue,
  and Woodcock}}]{BLW10}
\bibinfo{author}{\bibfnamefont{M.~N.} \bibnamefont{Bannerman}},
  \bibinfo{author}{\bibfnamefont{L.}~\bibnamefont{Lue}}, \bibnamefont{and}
  \bibinfo{author}{\bibfnamefont{L.~V.} \bibnamefont{Woodcock}},
  \bibinfo{journal}{J. Chem. Phys.} \textbf{\bibinfo{volume}{132}},
  \bibinfo{pages}{084507} (\bibinfo{year}{2010}).

\bibitem[{\citenamefont{Lab\'ik and Smith}(1994)}]{LS94}
\bibinfo{author}{\bibfnamefont{S.}~\bibnamefont{Lab\'ik}} \bibnamefont{and}
  \bibinfo{author}{\bibfnamefont{W.~R.} \bibnamefont{Smith}},
  \bibinfo{journal}{Mol. Simul.} \textbf{\bibinfo{volume}{12}},
  \bibinfo{pages}{23} (\bibinfo{year}{1994}).

\bibitem[{\citenamefont{Clisby and McCoy}(2006)}]{CM06}
\bibinfo{author}{\bibfnamefont{N.}~\bibnamefont{Clisby}} \bibnamefont{and}
  \bibinfo{author}{\bibfnamefont{B.~M.} \bibnamefont{McCoy}},
  \bibinfo{journal}{J. Stat. Phys.} \textbf{\bibinfo{volume}{122}},
  \bibinfo{pages}{15} (\bibinfo{year}{2006}).

\end{thebibliography}

\end{document}